\documentclass[11pt,a4paper]{article}
\usepackage{jcappub}
\usepackage{graphics}
\usepackage{dcolumn}
\usepackage{amsfonts}
\usepackage{amssymb}
\usepackage{bm}
\usepackage{hyperref}
\usepackage{url}
\usepackage{subfigure}
\usepackage[sort&compress]{natbib}
\usepackage{txfonts}
\usepackage{color}

\begin{document}
\newcommand{\changeR}[1]{\textcolor{red}{#1}}

\newcommand{\TBB}{{{T_{\rm BB}}}}
\newcommand{\TCMB}{{{T_{\rm CMB}}}}
\newcommand{\Te}{{{T_{\rm e}}}}
\newcommand{\Teq}{{{T^{\rm eq}_{\rm e}}}}
\newcommand{\Ti}{{{T_{\rm i}}}}
\newcommand{\nB}{{{n_{\rm B}}}}
\newcommand{\nHe}{{{n_{\rm ^4He}}}}
\newcommand{\nHet}{{{n_{\rm ^3He}}}}
\newcommand{\nHt}{{{n_{\rm { }^3H}}}}
\newcommand{\nHtw}{{{n_{\rm { }^2H}}}}
\newcommand{\nBes}{{{n_{\rm { }^7Be}}}}
\newcommand{\nLis}{{{n_{\rm { }^7Li}}}}
\newcommand{\nLisi}{{{n_{\rm { }^6Li}}}}
\newcommand{\nS}{{{n_{\rm s}}}}
\newcommand{\Teff}{{{T_{\rm eff}}}}

\newcommand{\id}{{{\rm d}}}
\newcommand{\aR}{{{a_{\rm R}}}}
\newcommand{\bR}{{{b_{\rm R}}}}
\newcommand{\neb}{{{n_{\rm eb}}}}
\newcommand{\neql}{{{n_{\rm eq}}}}
\newcommand{\kB}{{{k_{\rm B}}}}
\newcommand{\EB}{{{E_{\rm B}}}}
\newcommand{\zmin}{{{z_{\rm min}}}}
\newcommand{\zmax}{{{z_{\rm max}}}}
\newcommand{\YBEC}{{{Y_{\rm BEC}}}}
\newcommand{\YSZ}{{{Y_{\rm SZ}}}}
\newcommand{\rhob}{{{\rho_{\rm b}}}}
\newcommand{\Ne}{{{n_{\rm e}}}}
\newcommand{\sigT}{{{\sigma_{\rm T}}}}
\newcommand{\me}{{{m_{\rm e}}}}
\newcommand{\nBB}{{{n_{\rm BB}}}}

\newcommand{\kD}{{{{k_{\rm D}}}}}
\newcommand{\KC}{{{{K_{\rm C}}}}}
\newcommand{\KdC}{{{{K_{\rm dC}}}}}
\newcommand{\Kbr}{{{{K_{\rm br}}}}}
\newcommand{\zdC}{{{{z_{\rm dC}}}}}
\newcommand{\zbr}{{{{z_{\rm br}}}}}
\newcommand{\aC}{{{{a_{\rm C}}}}}
\newcommand{\adC}{{{{a_{\rm dC}}}}}
\newcommand{\abr}{{{{a_{\rm br}}}}}
\newcommand{\gdC}{{{{g_{\rm dC}}}}}
\newcommand{\gbr}{{{{g_{\rm br}}}}}
\newcommand{\gff}{{{{g_{\rm ff}}}}}
\newcommand{\xe}{{{{x_{\rm e}}}}}
\newcommand{\alphafs}{{{{\alpha_{\rm fs}}}}}
\newcommand{\YHe}{{{{Y_{\rm He}}}}}
\newcommand{\SE}{{{\dot{{\mathcal{E}}}}}}
\newcommand{\SQ}{{{{{\mathcal{E}}}}}}
\newcommand{\SN}{{\dot{\mathcal{N}}}}
\newcommand{\Sn}{{{\mathcal{N}}}}
\newcommand{\muc}{{{{\mu_{\rm c}}}}}
\newcommand{\xc}{{{{x_{\rm c}}}}}
\newcommand{\xH}{{{{x_{\rm H}}}}}
\newcommand{\mT}{{{{\mathcal{T}}}}}
\newcommand{\Ob}{{{{\Omega_{\rm b}}}}}
\newcommand{\Or}{{{{\Omega_{\rm r}}}}}
\newcommand{\Odm}{{{{\Omega_{\rm dm}}}}}
\newcommand{\mdm}{{{{m_{\rm WIMP}}}}}
\newcommand{\Hn}{\mathbf{\hat{n}}}
\newcommand{\npl}{{{n_{\rm Pl}}}}
\newcommand{\T}{T_0+T_1(\Hn)}

\title{Motion induced second order temperature and $y$-type anisotropies
  after the  subtraction of linear dipole in
  the CMB maps}

\author[a,b]{Rashid A. Sunyaev,}
\author[a]{Rishi Khatri}

\affiliation[a]{ Max Planck Institut f\"{u}r Astrophysik\\, Karl-Schwarzschild-Str. 1
  85741, Garching, Germany }
\affiliation[b]{Space Research Institute, Russian Academy of Sciences, Profsoyuznaya
 84/32, 117997 Moscow, Russia}
\date{\today}
\emailAdd{khatri@mpa-garching.mpg.de}
\abstract
{
$y$-type spectral distortions of the cosmic microwave background allow us
to detect clusters and groups of galaxies, filaments of hot gas and the
non-uniformities in the warm hot intergalactic medium. Several CMB
experiments (on small areas of sky) and theoretical groups (for full sky)
have recently published $y$-type distortion maps. We propose to search for
two artificial hot spots in such $y$-type maps resulting from the
incomplete subtraction of  the effect of the motion induced dipole on the cosmic microwave
background sky. This dipole introduces, at second order, additional
temperature and $y$-distortion anisotropy on the sky of amplitude few $\mu
K$ which could
potentially be measured by Planck HFI and Pixie experiments and can be used as a source of cross
channel calibration by CMB experiments. This $y$-type distortion is present in
every pixel and is not the result of averaging the whole sky. This
distortion, calculated  exactly from the known linear dipole, can be
subtracted from the final $y$-type maps, if desired.
}

\maketitle
\flushbottom
\section{Introduction}
Recently South Pole Telescope (SPT) \cite{spt} and Atacama Cosmology Telescope (ACT) \cite{act}
teams published  their power spectra for the thermal $y$-type distortions 
\cite{sz1980} for high $\ell$ multipoles. There are also many theoretical
computations of $y$-type distortion maps going down to $y\sim
  10^{-7}-10^{-6}$ \cite{tbo,bbps,lnb,ds2013}, where $y$ is the
amplitude of the $y$-type distortion, with
the expectation that Planck Surveyor's High Frequency Instrument (Planck-HFI) \cite{planck} will be able to
create maps of $y$-type distortion for the full sky.   Future proposed
experiments Cosmic Origins Explorer (CoRE) \cite{core}, Lite satellite for the studies of
B-mode polarization and Inflation from cosmic background Radiation Detection
 (LiteBIRD) \cite{litebird}  and Primordial
Inflation Explorer (Pixie) \cite{pixie} will be able to improve
the sensitivity by more than an order of magnitude over Planck-HFI and several
orders of magnitude compared to Cosmic Background Explorer's Far Infrared
Absolute Spectrophotometer  (COBE-FIRAS) \cite{firas}. Pixie will have an
angular resolution of approximately a  degree. The purpose of this short
note is to remind the observers about an additional artificial component
having a $y$-type spectrum which arises in the usual procedure of
subtracting the dipole component from  cosmic microwave background (CMB)
maps.

It is well known that sum of blackbody spectra is not a blackbody
\cite{zis1972} and that the averaging of dipole component of CMB over the
whole sky should lead to $y$-type distortion with amplitude $y\approx
2.5\times 10^{-7}$ \cite{cs2004,ksc2012b}. In this note we give the exact
formulae for the angular dependence and amplitude of the $y$-type distortions
in  CMB 
sky maps after the usual subtraction of the linear dipole component
described in the  Planck pre-launch and early results papers \cite{planck1,planck2}. It is important that the experimentalists will take special
search of two maxima  on the $y$-type distortion
maps peaked, one towards and the other exactly opposite, in  the direction of our motion with
respect to CMB  with the
characteristic dimensions of $38~{\rm deg}^2$ for 
$y>80\%$ of maximum and $105~{\rm deg}^2$ for $y>50\%$ of maximum. If  this
feature is discovered in $y$ maps, it would provide a source of
calibration for different channels of instruments such as Planck-HFI, in
addition to the solar dipole and the orbital dipole of the spacecraft. Relative
small angular dimensions of the hot spots open a possibility of their
detection by experiments like SPT and ACT or even experiments with much poorer
angular resolution. Planck-HFI, CoRE, LiteBIRD and Pixie are unique because they are
able to produce full sky $y$-type distortion  maps.

 In addition to the space missions Relikt \cite{strukov1987}, COBE
  \cite{cobedipole}  and Wilkinson Microwave
  Anisotropy Probe (WMAP) \cite{wmapdipole}, there have been numerous
  ground based and airborne measurements of the CMB dipole, see
  \cite{lineweaver1997} for a complete list.
 Since we know the dipole amplitude ($3.355\pm 0.008 ~{\rm mK}$) and
 direction (galactic longitude $263.99\pm0.14$ deg, latitude $48.26\pm
 0.03$ deg) at high precision from COBE Differential Microwave Radiometers
 (DMR) and  (WMAP) measurements \cite{cobedipole,wmapdipole}, these
artificial 
second order temperature and $y$-type quadrupoles can be easily subtracted
from the CMB maps. 

It is important to mention that the  evaluation of the terms of second
order in $\beta=v/c$, where $v$ is our motion w.r.t. CMB and $c$ is the
speed of light, gives us an additional term with blackbody spectrum 
but
with characteristic quadrupolar angular dependence over the sky in the large scale CMB
maps. This term also has two positive maxima coinciding with the axes of
dipole and the $y$-type distortion  is comparable in magnitude to this well
known second order quadrupole effect. The blackbody part of the motion induced quadrupole will add to the
 primary CMB quadrupole, but both these components would be absent in the
 $y$-type distortion maps.

This frequency dependence of the
second order terms in the Taylor series expansion of motion induced
anisotropies in the CMB was  studied
by several authors \cite{sz1980b,deg1990, bdm1992,ss1999,kk2003,cs2004}. However, the connection with
the $y$-type distortion,  in particular the possibility of the residual
$y$-type anisotropy due to our motion in the full sky  $y$-type maps
produced by highly sensitive experiments like Planck-HFI, was not made in the
previous studies. See also \cite{kk2003} on additional aspects of the
motion induced quadrupole not discussed here.

\section{y-type distortion and temperature anisotropy in CMB maps from
  motion induced dipole}
In each direction $\Hn$ in the sky, CMB consists of a blackbody with temperature
$T_0+T_1(\Hn)$, where $T_0\approx 2.725~\rm{K}$ is the average temperature
and $T_0+T_1(\Hn)\equiv T_0+T_1(\theta)=T_0(1+\beta
\cos(\theta))^{-1}(1-\beta^2)^{1/2}$ \cite{ll1971,pw1968}  is the velocity induced
anisotropy, where $\beta=1.23\times 10^{-3}\pm 0.2\%$ is the velocity of sun w.r.t.
CMB in units of speed of light in the direction
$\Hn_1=(\ell=263.99,b=48.26)$ in galactic coordinates \cite{cobedipole,wmapdipole} and
$\cos(\theta)=\Hn.\Hn_1$. Taylor expanding to
second order in $\beta$, we get for the velocity induced monopole, dipole and
quadrupole,  $T_1(\theta)\approx T_0 \left[\beta \cos
(\theta)+\beta^2\left(\cos^2(\theta)-1/2\right)\right]$. Therefore the
intensity or equivalently the occupation number ($I_{\nu}=2h\nu^3/c^2 n(\nu)$) in each direction in
the sky is given by, following \cite{sz1980b},
\begin{align}
&\npl\left(T_0+T_1(\Hn)\right)\equiv \frac{1}{e^{\frac{h\nu}{\kB\left(\T\right)}}-1}\nonumber\\
\approx &\npl(T_0)+\ln\left[1+\frac{ T_1(\Hn)}{T_0}\right]\frac{\partial \npl(T_0)}{\partial
  \ln[T_0]}+\frac{1}{2}\left(\ln\left[1+\frac{T_1(\Hn)}{T_0}\right]\right)^2\frac{\partial^2 \npl(T_0)}{\partial
  (\ln[T_0])^2}\nonumber\\
=&\npl(T_0)+\left(\frac{T_1(\Hn)}{T_0}+\left(\frac{T_1(\Hn)}{T_0}\right)^2\right)T_0\frac{\partial \npl(T_0)}{\partial
  T_0}+\frac{1}{2}\left(\frac{T_1(\Hn)}{T_0}\right)^2T_0^4\frac{\partial }{\partial
  T_0}\frac{1}{T_0^2}\frac{\partial \npl(T_0)}{\partial T_0}\nonumber\\
=&\npl(T_0)+\left(\frac{T_1(\Hn)}{T_0}+\left(\frac{T_1(\Hn)}{T_0}\right)^2\right)T_0\frac{\partial \npl(T_0)}{\partial
  T_0}+\frac{1}{2}Y(x)\left(\frac{T_1(\Hn)}{T_0}\right)^2,\label{taylor}
\end{align}
where $h$ is the Planck's
constant and $\kB$ is the Boltzmann's constant.
Planck-HFI is insensitive to the average intensity and only measures the
fluctuating part,
$\npl\left(T_0+T_1(\Hn)\right)-\npl\left(T_0\right)$. We are
therefore subtracting, and mixing,  two blackbodies of different
temperature which should give us a $y$-type distortion $+$ change in the
temperature \cite{zis1972,cs2004,ksc2012b}. 

\begin{align}
\frac{c^2}{2h\nu^3}\Delta I_{\nu}&=\left(\frac{T_1(\Hn)}{T_0}+\left(\frac{T_1(\Hn)}{T_0}\right)^2\right)G(x)+\frac{1}{2}Y(x)\left(\frac{T_1(\Hn)}{T_0}\right)^2\nonumber\\
&=\left[\beta \cos(\theta) +\beta^2\left(2\cos^2(\theta)-\frac{1}{2}\right)\right]G(x)+\frac{1}{2}Y(x)\left(\beta\cos(\theta)\right)^2,
\end{align}
where $x=h\nu/\kB T_0$ is the dimension less frequency, 
\begin{align}
G(x)&=\frac{xe^x}{(e^x-1)^2}\nonumber\\
&\rm{and}\nonumber\\
Y(x)&=\frac{xe^x}{(e^x-1)^2}\left(x\frac{e^x+1}{e^x-1}-4\right)
\end{align}
is the $y$-type distortion.

We can also subtract, and absorb in $T_0$, an average term, $(\beta^2/6)G(x)$ to make the $\beta^2G(x)$ term proportional to
the Legendre polynomial $P_2(\cos(\theta))$. Similarly, we should also subtract
average $y$-type distortion.
Subtracting the linear dipole thus leaves the following residual in the map,
\begin{align}
\frac{c^2}{2h\nu^3}I_{\nu}^{\rm residual}=\beta^2\left(2\cos^2(\theta)-\frac{2}{3}\right)G(x)+\frac{1}{2}Y(x)\beta^2\left(\cos^2(\theta)-\frac{1}{3}\right)\label{inu}
\end{align}
The first term is just change in the  blackbody temperature,  the second term is the   $y$-distortion
amplitude and might become visible in the future y-distortion maps of Pixie
\cite{pixie}  and CoRE \cite{core}, and possibly in the Planck-HFI \cite{planck} maps. Pixie will also
be sensitive to the monopole, $y=\beta^2/6=2.5\times 10^{-7}$. This average
distortion is larger than what is expected from reionization \cite{cs2004,pixie}. Note that after 
subtracting the monopole, the quadrupolar $y$-type distortion parameter is negative 
in part of the sky with minimum around $\theta=\pi/2, 3\pi/2$.

\begin{figure}
\includegraphics[width=14cm]{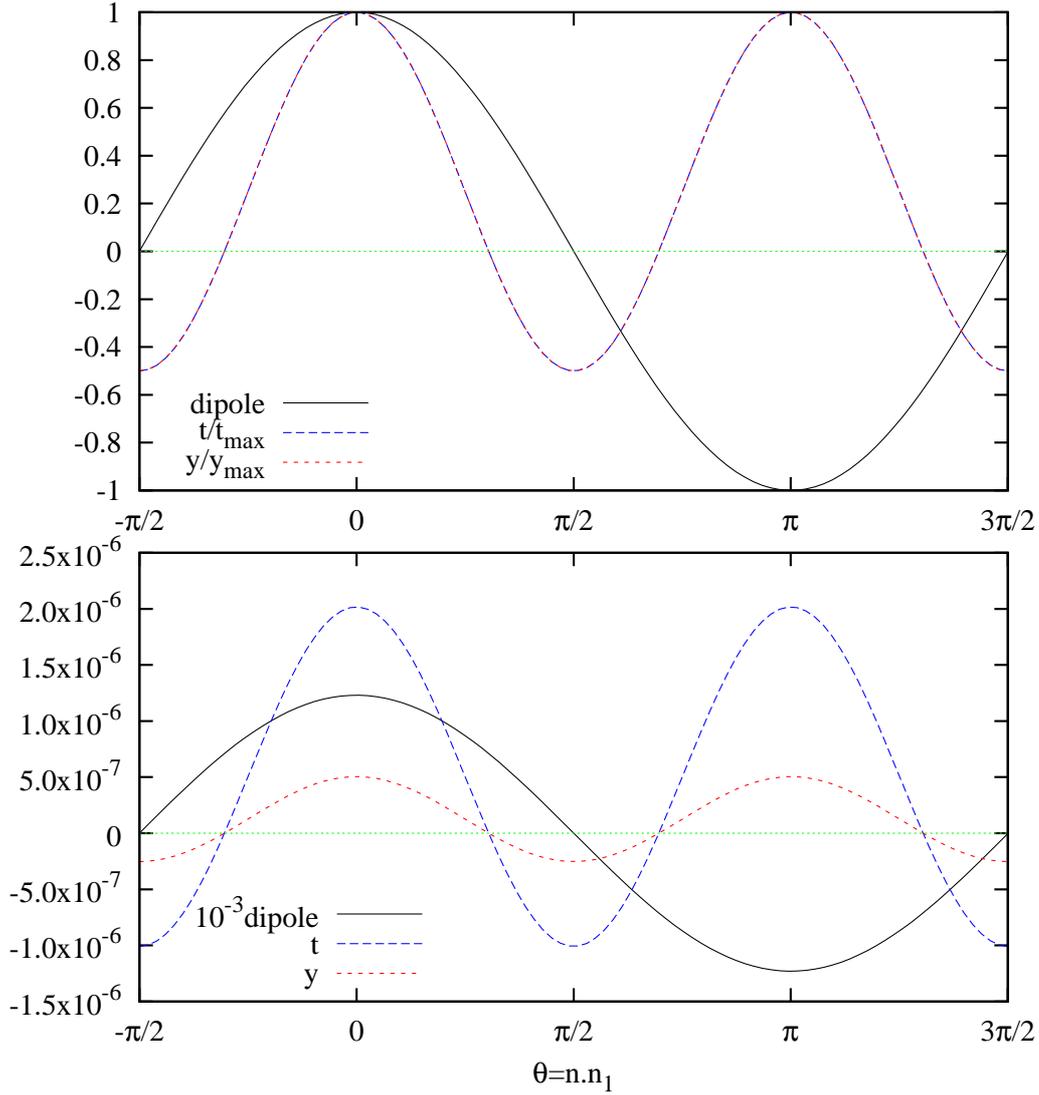}
\caption{\label{dipolefig}{Temperature and $y$-distortion anisotropy as a
    function of angle from the CMB dipole axes, $\theta\equiv
    \Hn.\Hn_1$. Upper panel shows the angular distribution of dipole, temperature and $y$-type
    anisotropies relative to their respective maxima while the bottom panel shows
    their actual values. }}
\end{figure}

\begin{figure}
\includegraphics[width=14cm]{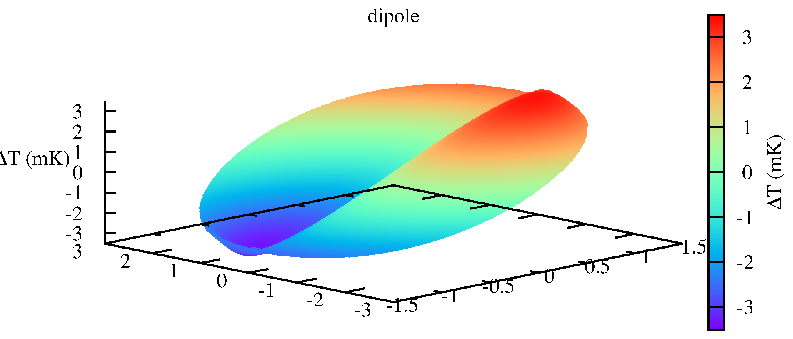}
\includegraphics[width=14cm]{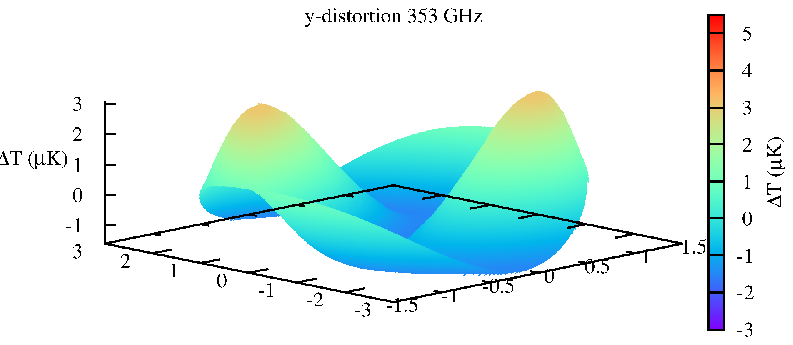}
\includegraphics[width=14cm]{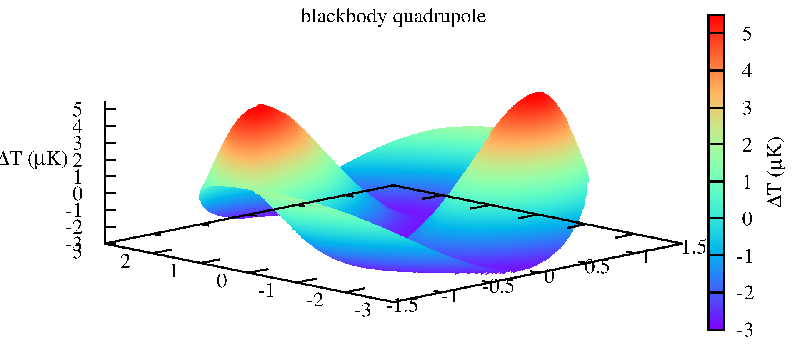}
\caption{\label{dipole3d}{Full sky maps in equal-area Mollweide projection
   on the x-y plane for the dipole, second order motion induced blackbody temperature
   quadrupolar anisotropy and motion induced $y$-type quadrupolar anisotropy
 is shown. Only the $y$-type quadrupolar component would be present in the
 $y$-type distortion maps produced by CMB experiments such as Planck-HFI.}}
\end{figure}

The maximum residual occurs at $\theta=0,\pi$ and is given by, in CMB
temperature units,  
\begin{align}\frac{c^2}{2h\nu^3} 
\frac{I_{\nu}^{\rm residual}T_0}{G(x)} = T_0\left[\frac{4}{3}\beta^2+\frac{1}{3}\beta^2Y(x)/G(x)\right].
\end{align}
The maximum  change in the  blackbody temperature is
  $T_0t_{\rm max}=T_0\frac{4}{3}\beta^2=5.5~{\rm \mu K}$. The maximum $y$-distortion
amplitude is $y_{\rm max}=\frac{1}{3}\beta^2=5\times 10^{-7} \pm
0.5\%$. The uncertainities in these values depend only on our knowledge of
the dipole which is known with an accuracy of $8 ~\mu{\rm K}$  \cite{wmapdipole}. Pixie will have absolute sensitivity
of few nK \cite{pixie} and the $y$-distortion quadrupole, also known with
a similar precision, might be an
important external source of calibration for it. Table \ref{tbl1} gives
the maximum signal for different Planck-HFI and Low Frequency Instrument
(LFI) channels. Figure \ref{dipolefig}
shows the linear dipole, additional blackbody temperature anisotropy $t=
\beta^2\left(2\cos^2(\theta)-\frac{2}{3}\right)$ and $y$-type anisotropy
$y=\frac{1}{2}\beta^2\left(\cos^2(\theta)-\frac{1}{3}\right)$ and their relative
amplitude with respect their maxima as a function of angle $\theta$ of the
line of sight direction $\Hn$ with the dipole $\Hn_1$. Figure \ref{dipole3d} shows the 3d visualization of the dipole,  and second
order temperature and $y$-distortion (353 GHz) anisotropies with Mollweide
equal area projection on the x-y plane and amplitude in color scale as well
as the z axes. The temperature quadrupole ($\beta^2 G(x)$ term) in particular has a different
orientation than the primary CMB quadrupole \cite{wmapq} but much smaller
amplitude, $Q_{\rm rms}^t=(5/(4\pi)C_{\ell=2})^{1/2}=4\beta^2/(3\sqrt{5})T_0= 2.5 \rm \mu{\rm K}$.
The area on the sky
covered by a spot with $y>0.8 y_{\rm max}$ is $38 ~{\rm deg}^2$ and the
region with $y>0.5 y_{\rm max}$ covers $105 ~{\rm deg}^2$. Thus even though
 this signal may be not be detectable in individual pixels, the peaks,
 with more than a $1000$ pixels, might be detected.  This $y$-type
distortion is artificial, resulting from incomplete subtraction of the
effect of our motion from the CMB maps, but it still opens a way to
calibrate CMB experiments, as was already noted by
Refs. \cite{kk2003,cs2004} but Ref. \cite{kk2003}
 did not separate out $y$-type part.

\begin{table}
\begin{tabular}{|c|c|c|c|}
\hline
 Channel  & $\Delta T=T_0\frac{4}{3}\beta^2$ ($\mu\rm{K}$) & y-distortion ($\mu\rm{K}$) & total ($\mu\rm{K}$)\\
\hline
30 & & -2.7 & 2.8\\
44 & & -2.6 & 2.9\\
70 & & -2.4 & 3.1\\
100 &5.5 & -2.1 & 3.4\\
143 & & -1.4 & 4.1\\
217 & & 0.0 & 5.5\\
353 & & 3.1 & 8.6\\
\hline
\end{tabular}
\caption{\label{tbl1}Maximum residuals in CMB temperature units in
  Planck-HFI and LFI channels after removing
  the linear dipole. The blackbody temperature part is absent from the $y$-type distortion maps produced by SPT,ACT,Planck-HFI.} 
\end{table}

\bibliographystyle{JHEP}
\bibliography{dipole}
\end{document}